\documentclass[nofootinbib,superscriptaddress,prl,twocolumn]{revtex4-1}
\usepackage{amsmath,graphicx,hyperref}
\allowdisplaybreaks

\newcommand{\mc}{\mathcal}
\newcommand{\mr}{\mathrm}

\newcommand{\be}{\begin{equation}} 
\newcommand{\ee}{\end{equation}} 
\newcommand{\bea}{\begin{eqnarray}} 
\newcommand{\eea}{\end{eqnarray}}

\newcommand{\n}{\overline{n}}

\newcommand{\nnb}{\nonumber}

\newcommand{\ec}{E_{\mr{cut}}}
\newcommand{\cs}{\overline{cs}}

\begin{document}



\title{Universal lepton universality violation in exclusive processes}

\def\Seoultech{Institute of Convergence Fundamental Studies and School of Liberal Arts, Seoul National University of Science and Technology, Seoul 01811, Korea}
\def\Pitt{Pittsburgh Particle Physics Astrophysics and Cosmology Center (PITT PACC) \\ Department of Physics and Astronomy, University of Pittsburgh, Pittsburgh, Pennsylvania 15260, USA}
\def\BenGurion{Department of Physics, Ben-Gurion University of the Negev, Beer-Sheva 84105, Israel}

\author{Lin Dai}
\email[E-mail:]{dail@post.bgu.ac.il}
\affiliation{\BenGurion}
\author{Chul Kim}
\email[E-mail:]{chul@seoultech.ac.kr}
\affiliation{\Seoultech} 
\author{Adam K. Leibovich}
\email[E-mail:]{akl2@pitt.edu}
\affiliation{\Pitt}

\begin{abstract} 
In high energy exclusive processes involving leptons, QED corrections can be sensitive to infrared scales like the lepton mass and the soft photon energy cut, resulting in large logarithms that need to be resummed to all order in $\alpha$.
When considering the ratio of the exclusive processes between two different lepton flavors, the ratio $R$ can be expressed in terms of factorized functions in the decoupled leptonic sectors. While some of the functional terms cancel, there remain the large logarithms due to the lepton mass difference and the energy cut. 
This factorization process can be universally applied to the exclusive processes such as $Z\to l^+l^-$ and $B^-\to l^-\bar{\nu}_l$, where the resummed result in the ratio gives significant deviations from the naive expectation from lepton universality. 
\end{abstract}

\maketitle 


Lepton universality due to the identical gauge coupling of the three families of lepton has played a key role in verifying the structure of the Standard Model (SM). 
Even though the mass gap between the different flavors is large, we may expect almost the same scattering cross sections or branching fractions in very high energy processes, where the lepton mass dependence can be safely ignored. The most representative example is inclusive $Z$ decays to a lepton pair. With the decay ratios between two different lepton flavors defined  by
\be 
\label{Rin} 
R_{12} = \frac{\Gamma(Z\to l_1\bar{l}_1 X)}{\Gamma(Z\to l_2\bar{l}_2 X)} = 1+ \Delta_{12}, 
\ee
the asymmetries have been measured and are known to be less than one percent, i.e., $|\Delta_{12}| < 1\%$~\cite{ALEPH:2005ab,Aaboud:2016btc}.

Lepton universality has also been tested in $B$ decays, where flavor transitions are highly entangled. (We refer to Ref.~\cite{Bifani:2018zmi} for the recent review.) 
Since the masses of electrons and muons are negligible compared to the $b$ quark mass, it is natural to expect that the decay modes with the electron and with the muon are almost identical in the SM.  
However, recently reported experimental data for  exclusive semileptonic $B$ decay, for example $B \to K^{(*)}l^+l^-$, show large asymmetries, even though they still include large errors~\cite{Lees:2012tva,Aaij:2014ora,Aaij:2017vbb,Aaij:2019wad}. 

Unlike for inclusive processes, in exclusive processes rather large differences in QED corrections can be expected between  lepton flavors even though the lepton masses are much smaller than the relevant decay or scattering hard scale. When the exclusive process is sensitive to infrared (IR) divergences, the cross section or the branching fraction for the process can be highly dependent on these IR scales~\cite{Bordone:2016gaq}, leading to disparate corrections.

In exclusive processes a photon energy cut is required experimentally, and it theoretically forbids soft IR divergences in QED corrections. The regularized soft divergence becomes a large logarithm of the soft photon energy cut. 
Although the fine structure constant $\alpha\sim 1/137$ at the IR scale is small, when combined with the large logarithm it can give a non-negligible QED contribution. The significance of the soft photon corrections has been considered in exclusive $B$ decay processes, for example $B\to Dl\bar{\nu}_l$~\cite{deBoer:2018ipi}, $B\to l \nu_l$~\cite{Becirevic:2009aq} and $B\to l^+l^-$ decays~\cite{Buras:2012ru,Beneke:2019slt}.    

The collinear IR divergences do not cancel in exclusive processes. Considering an energetic lepton in the final state, the (virtually) radiated photon parallel to the lepton gives rise to a collinear divergence. This divergence can be regularized using a non-zero lepton mass resulting in a large logarithm with the small mass in QED corrections. This logarithm could be the main source of lepton universality violation in SM predictions. In general the large logarithms from the soft and the collinear divergences are not fully separated, but instead  overlap to form Sudakov logarithms. Thus, even at one loop order, we face double logarithms, which should be resummed to all orders in $\alpha$.    

Recently, in the QCD factorization framework using soft-collinear effective theory (SCET)~\cite{Bauer:2000ew,Bauer:2000yr,Bauer:2001yt}, 
the exclusive heavy quark pair production in $e^+e^-$ annihilation was analyzed~\cite{Kim:2020dgu}.  
When the invariant mass of the heavy quark pair, $Q$, is much larger than the quark mass $m$, the quarks are highly boosted moving in opposite directions. Hence each heavy quark sector safely decouples from the hard interactions governed by the scale $Q$. In order to obtain an IR safe cross section,  an upper limit on the soft hadron's energy, i.e., the energy cut $\ec$, was imposed. Following the usual SCET factorization, the soft part of the gluon field for soft gluon radiations with $p_s \sim \ec$ was introduced. With this factorization, 
the large logarithms of $Q/m$ and $Q/\ec$ were resummed and it was found that the resummation dramatically changed the tree level result~\cite{Kim:2020dgu}. In this paper, we follow a similar analysis in the exclusive leptonic sector and will see this can be the main source for lepton universality violation. 

We begin by considering the exclusive $Z$ boson decay to a lepton pair with a small photon energy cut, $\ec$. 
This process describes an analogous situation to the heavy quark pair production mentioned above. We can therefore employ the same factorization theorem in Ref.~\cite{Kim:2020dgu}, converting QCD to QED interactions. The partial decay width can be written immediately as  
\bea 
\label{factZ}
\Gamma(Z\to l^+l^-) &=& \Gamma_0\cdot H(Q,\mu) S(\ec,\mu)  \\
&&~\times [C (m,\mu) B  (\beta m, \mu) ]^2, \nnb
\eea 
where $Q = m_Z$ and $\Gamma_0$ is the Born level partial decay width. $H$ is the hard function obtained from integrating out the hard interactions with offshellness $p_H^2 \sim Q^2$, and the soft function $S$ describes soft photon radiations limited by $\ec$. The functions $C$ and $B$ are described below. 

For this process, the leptons are highly energetic and are moving in opposite directions in the $Z$ boson rest frame. 
Thus, the leptons can be described by $n$- and $\n$-collinear interactions, where $n$ and $\n$ are the lightcone vectors satisfying $n^2=\n^2 =0$ and $n\cdot \n =2$. These collinear interactions have offshellness $\sim m^2$, and scale as 
\bea 
p_n^{\mu} &=&  (\n\cdot p_n, p_n^{\perp},n\cdot p_n) \equiv (p_n^+,p_n^{\perp},p_n^-) \sim Q(1,\lambda,\lambda^2) ,\nnb \\
p_{\n}^{\mu} &=&  (p_{\n}^+, p_{\n}^{\perp},p_{\n}^-) \sim Q(\lambda^2,\lambda,1), 
\eea
with $\lambda \sim m/Q$.  
A collinear photon would have large energy of order $Q$, so real radiation is forbidden and only virtual collinear radiation can be present. However, if the collinear photon becomes soft with the energy comparable with $\ec$, this collinear-soft (csoft) photon 
can lead to real radiation, with the contribution dependent upon $\ec$. These csoft interactions in the $n$ and $\n$ directions scale as 
\be
p_{cs}^{\mu}  \sim \ec(1,\lambda,\lambda^2),~~ 
p_{\cs}^{\mu}  \sim \ec (\lambda^2,\lambda,1). 
\ee
These csoft interactions have much smaller offshellness compared to the collinear interactions, $p_{cs}^2 \sim p_{\cs}^2 \sim \beta^2 m^2 \ll m^2$, where $\beta = \ec/Q$.  

The interactions of the leptonic sector can be separated into the collinear and the csoft interactions.  
To obtain this factorization, we first integrate out the collinear interactions, which gives the collinear function $C(m,\mu)$ in Eq.~\eqref{factZ}. Then the remaining and residual csoft interactions can be described in the QED version of the boosted heavy quark effective theory (bHQET)~\cite{Fleming:2007qr}. This contribution gives the csoft function $B(\beta m,\mu)$ in Eq.~\eqref{factZ}.

Each factorized function in Eq.~\eqref{factZ} is normalized to unity at leading order (LO) in $\alpha$.
The next-to-leading order (NLO) contributions are 
\begin{align}
H^{(1)}(Q,\mu) &= \frac{\alpha}{2\pi}\Bigl(-3\ln\frac{\mu^2}{Q^2} - \ln^2\frac{\mu^2}{Q^2} - 8+ \frac{7\pi^2}{6}\Bigr), \nnb \\
\label{fNLO}
S^{(1)}(\beta Q, \mu) &= \frac{\alpha}{2\pi} \Bigl(  \ln^2 \frac{\mu^2}{4\beta^2 Q^2} -\frac{\pi^2}{2} \Bigr), \\
C^{(1)}(m,\mu) &=  \frac{\alpha}{2\pi} \Bigl(\frac{1}{2} \ln\frac{\mu^2}{m^2}+\frac{1}{2} \ln^2 \frac{\mu^2}{m^2} +2+\frac{\pi^2}{12} \Bigr), \nnb \\
B^{(1)}(\beta m, \mu) &= \frac{\alpha}{2\pi} \Bigl(\ln \frac{\mu^2}{4\beta^2 m^2} - \frac{1}{2} \ln^2 \frac{\mu^2}{4\beta^2 m^2} -\frac{\pi^2}{12} \Bigr). \nnb
\end{align}
Note that all the factorized functions involve large double logarithms at one loop. Hence the decay width in Eq.~\eqref{factZ} is very sensitive to QED corrections, and the large logarithms need to be resummed to all orders in $\alpha$. 

The resummation of the large logarithms can be systematically performed using this factorization. 
From Eq.~\eqref{fNLO}, we can easily determine the characteristic scales that minimize the large logarithms in the factorized functions. They are given by 
\bea 
\mu_H \sim Q,~~\mu_S \sim 2\beta Q,~~\mu_C \sim m,~~\mu_B \sim 2\beta m. 
\eea 
We can then resum the large logarithms in the exclusive $Z$ decay through renormalization group evolution of the factorized functions from the characteristic scales to the factorization scale in Eq.~\eqref{factZ}.  

The factorization theorem in Eq.~\eqref{factZ} also has an advantage when we consider the ratio between the decays involving different lepton flavors. 
In the ratio, with the same energy cut applied, the hard and the soft functions exactly cancel, and the ratio is given by
\be
\label{rZ} 
R^Z_{12} \equiv \frac{\Gamma(Z\to l_1^+ l_1^-)}{\Gamma(Z\to l_2^+ l_2^-)} 
= \frac{[C (m_1,\mu) B  (\beta m_1, \mu) ]^2}{[C (m_2,\mu) B  (\beta m_2, \mu) ]^2}.
\ee
This ratio is sensitive to the QED corrections since the large logarithms sensitive to the masses of the two flavors are still present.

Let us consider the resummation of Eq.~\eqref{rZ} to next-to-leading logarithmic (NLL) accuracy, which resums all powers of $\alpha L$ and $\alpha L^2$, where $L$ denotes a large logarithm power-counted as $\mc{O}(1/\alpha)$.  
To achieve the resummation, $C$ and $B$ are evolved from their characteristic scales to the factorization scale in Eq.~\eqref{rZ} using the following anomalous dimensions:
\bea 
\label{anom}
\gamma_C &=& \frac{1}{C}\frac{dC}{d\ln\mu} =\Gamma_C (\alpha)  \ln\frac{\mu^2}{m^2} + \frac{\alpha}{2\pi}, \\ 
\gamma_B &=& \frac{1}{B}\frac{dB}{d\ln\mu}= -\Gamma_C (\alpha)  \ln\frac{\mu^2}{4\beta^2m^2} + \frac{\alpha}{\pi}. \nnb
\eea 
Here the QED cusp anomalous dimension, $\Gamma_C$, can be expanded as $\Gamma_C=\sum_{k=0} \Gamma_k (\alpha/4\pi)^{k+1}$. To NLL accuracy, we need the first two coefficients~\cite{Billis:2019evv}, 
\be 
\Gamma_0 = 4,~~\Gamma_1 = 4\Bigl[-\frac{20}{9} \Bigl(3 \sum_q Q_q^2 + n_l\Bigr)\Bigr],
\ee
where $Q_q$ is the electric charge of the quark $q$ and $n_l$ is the number of the lepton flavors. 

\begin{figure*}[t]
\begin{center}
\includegraphics[width=16cm]{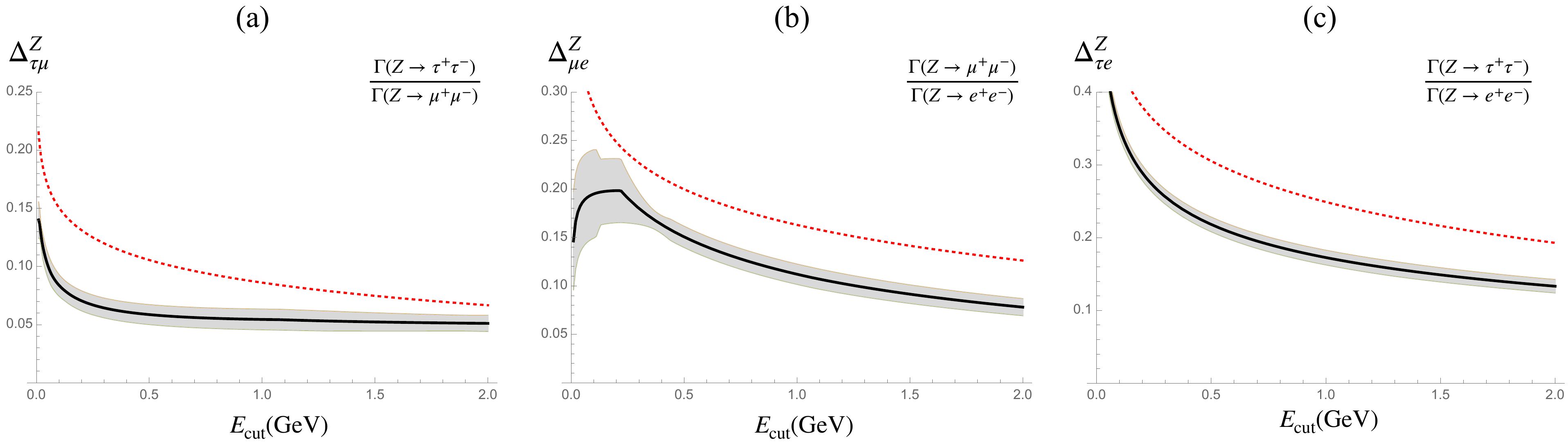}
\end{center}
\vspace{-0.7cm}
\caption{\label{fig1} 
Lepton universality violations, $\Delta^Z_{ij} = R_{ij}^Z -1~(i,j=\tau,\mu,e)$, in exclusive leptonic $Z$ boson decays. Here the red dotted lines denote the fixed-order NLO results, and the black solid lines with gray bands are the resummed results at NLL accuracy with associated errors from varying the characteristic scales.    
 } 
\end{figure*}

Evolving the $C$s and $B$s in Eq.~\eqref{rZ} using the anomalous dimensions in Eq.~\eqref{anom}, we resum the large logarithms with NLL accuracy. 
Since $\mc{O}(\alpha)$ corrections are ignored at NLL, the resummed result can be expressed in exponential form,
\begin{widetext} 
\bea 
\label{resR} 
\ln R_{12}^Z &=& -4 S_{\Gamma} (\mu_{1C},\mu_{2C}) + 4 S_{\Gamma} (\mu_{1B},\mu_{2B})  +2\ln\frac{m_1^2}{m_2^2} a_{\Gamma} (\mu_{2C},\mu_{2B})  \\
&& - 2\ln\frac{\mu_{1C}^2}{m_1^2} a_{\Gamma} (\mu_{1C},\mu_{2C}) +2 \ln\frac{\mu_{1B}^2}{4\beta^2 m_1^2} a_{\Gamma} (\mu_{1B},\mu_{2B})
-\left(\int^{\mu_{C1}}_{\mu_{C2}}\frac{d\mu}{\mu}+2\int^{\mu_{B1}}_{\mu_{B2}}\frac{d\mu}{\mu}\right)\frac{\alpha(\mu)}{\pi}\ . \nnb 
\eea 
\end{widetext}
Here the evolution functions $S_{\Gamma}$ and $a_{\Gamma}$ are given by 
\bea 
S_{\Gamma}(\mu_1,\mu_2) &=& \int^{\mu_1}_{\mu_2} \frac{d\mu}{\mu} \Gamma_C(\alpha) \ln\frac{\mu}{\mu_1}  \\
&=& \int^{\alpha_1}_{\alpha_2} \frac{d\alpha}{b_e(\alpha)} \Gamma_C(\alpha) 
\int^{\alpha}_{\alpha_1} \frac{d\alpha'}{b_e(\alpha')}\ ,  \nnb \\
a_{\Gamma}(\mu_1,\mu_2) &=& \int^{\mu_1}_{\mu_2} \frac{d\mu}{\mu} \Gamma_C(\alpha)= \int^{\alpha_1}_{\alpha_2} \frac{d\alpha}{b_e(\alpha)} \Gamma_C(\alpha), \nnb 
\eea
where $\alpha_{1,2} \equiv \alpha(\mu_{1,2})$ and $b_e(\alpha)$ is the QED beta function given by $d\alpha/d\ln\mu$. 
On the right-hand side of Eq.~\eqref{resR}, the terms in the first line are the resummed result of the double logarithms, which gives contributions of $\mc{O}(1/\alpha)$, and the remaining terms are the contributions of $\mc{O}(1)$.

For illustration purposes, keeping the coupling $\alpha$ constant under scale variance, 
we approximate Eq.~\eqref{resR} as  
\be
\label{Rfix} 
R^Z_{12} \approx \exp \Bigl[-4\Gamma_C (\alpha) \ln \frac{m_1}{m_2} \ln 2\beta -\frac{3\alpha}{\pi} \ln\frac{m_1}{m_2}\Bigr],   
\ee
where we have set $\mu_{iC} = m_{i},~\mu_{iB}=2\beta m_i~(i=1,2)$. If we expand Eq.~\eqref{Rfix} to order $\alpha$, we recover the fixed-order NLO result of Eq.~\eqref{rZ} such as 
\be
\label{RNLO} 
R^Z_{12,\rm{NLO}} = 1- \frac{\alpha}{\pi} \Bigl[4 \ln \frac{m_1}{m_2} \ln 2\beta+3\ln\frac{m_1}{m_2}\Bigr].
\ee

In FIG.~\ref{fig1}, we illustrate the resummed results (black solid lines) for the lepton asymmetries, $\Delta^Z_{ij} = R_{ij}^Z -1~(i,j=\tau,\mu,e)$, between the exclusive leptonic decays of $Z$ boson, varying the photon energy cut. The errors have been estimated by varying each characteristic scale $\mu_i$ from $\mu_i^0/2$ to $2\mu_i^0$ independently, where the default characteristic scales have been chosen to be $\mu_{iC}^0 = m_{i},~\mu_{iB}^0=2\beta m_i~(i=\tau,\mu,e)$. 

The heavier the lepton, the larger the decay width becomes. Hence the asymmetries in FIG.~\ref{fig1} consistently give positive numbers. 
When compared with the fixed-order NLO results, the sizes of the asymmetries in the resummed results are reduced, but still give significant deviations from zero. For example, for $\ec = 1~\mr{GeV}$, the asymmetries in the resummed results are given by 
\bea
\Delta^Z_{\tau\mu} &=& 0.054\pm 0.009,~~\Delta^Z_{\mu e}=0.112\pm0.010, \nnb \\
\Delta^Z_{\tau e} &=& 0.173\pm 0.010. 
\eea
Note that these asymmetries for the exclusive processes are considerably enhanced when compared with ones for the inclusive processes shown in Eq.~\eqref{Rin}. The reason is that the exclusive processes have the large logarithmic dependencies on the small lepton mass and $\ec$ while the inclusive processes do not. 
As far as the energy cut parameter $\beta$ is given to be small, we can apply the resummed result in Eq.~(\ref{resR}) to the exclusive heavy vector meson decays, for example $\Upsilon \to l^+l^-$ or $J/\Psi \to l^+l^-$, where QCD contributions inside the meson cancel in the ratio of the decay widths with different lepton flavors.

We can also estimate the lepton universality violation in the exclusive leptonic $B$ decays, $B^- \to l^-\bar{\nu}_l$, from the resummed result.\footnote{
We may consider the violation effect on the processes $B\to l^+l^-$ with a similar reasoning. However, the presence of the power-enhanced contributions by $m_b/\Lambda_{QCD}$ in QED corrections~\cite{Beneke:2017vpq,Beneke:2019slt} gives rise to additional contributions to the violation and complicates the factorization. 
} 
Since the $B$ meson is a pseudoscalar, the partial decay width is proportional to the lepton mass squared, and the ratio between  different lepton flavors is given by
\bea
\label{RB12}
R_{12}^B &=& \frac{m_1^2}{m_2^2} \Bigl(\frac{m_B^2 - m_1^2}{m_B^2 - m_2^2}\Bigr)^2 \frac{C (m_1,\mu) B (\beta m_1, \mu)}{C(m_2,\mu) B  (\beta m_2, \mu)} \\
&=&  \frac{m_1^2}{m_2^2} \Bigl(\frac{m_B^2 - m_1^2}{m_B^2 - m_2^2}\Bigr)^2 (1+ \Delta^B_{12}). \nnb 
\eea   
Here the contributions from strong and weak interactions cancel. The remaining contributions are the QED corrections to the charged leptonic sector, which can also be factorized into $C$ and $B$ in the heavy $b$ quark limit. 
Since we have a single charged lepton in this process, the exponentiation factor from the resummation takes a half of the contribution to the exclusive $Z$ decays, i.e, it is given by $(\ln R_{12}^Z)/2|_{m_Z \to m_B}$.

\begin{figure}[h]
\begin{center}
\includegraphics[height=5.1cm]{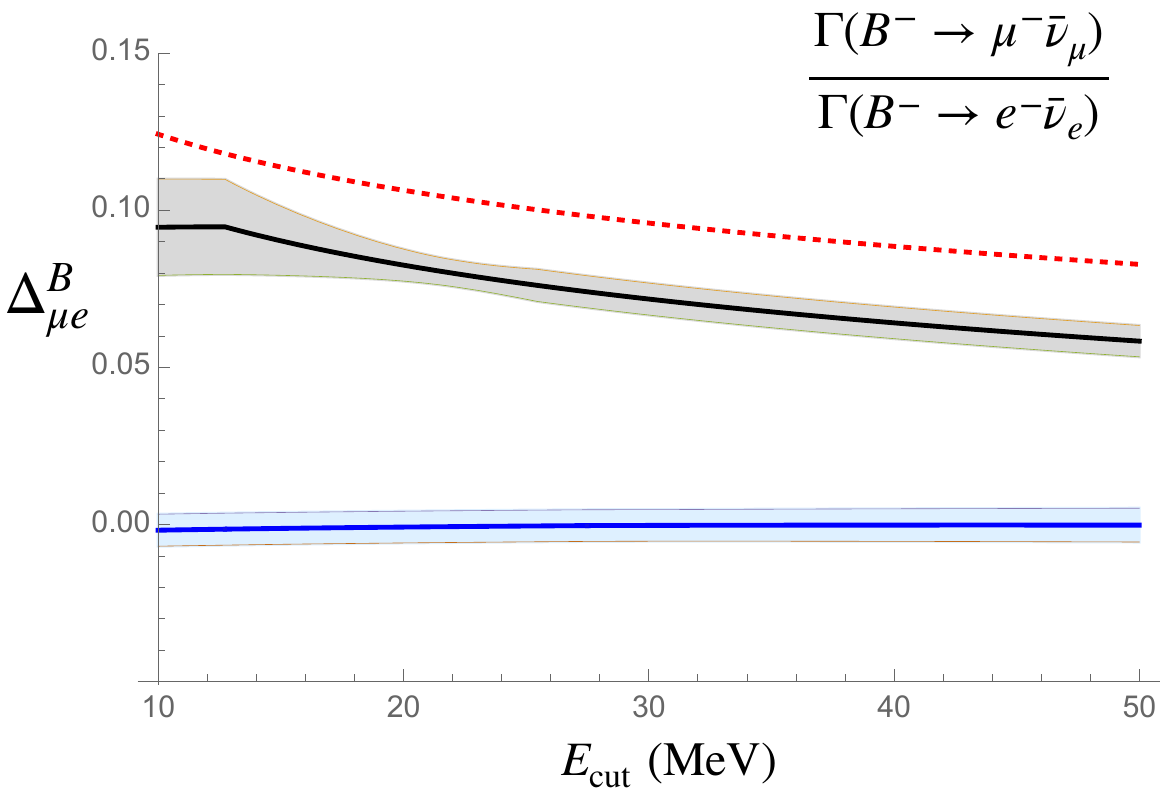}
\end{center}
\vspace{-0.7cm}
\caption{\label{fig2} 
Lepton universality violation in $B^- \to l^-\bar{\nu}_l~(l=\mu,e)$. The red-dotted line is the fixed-order NLO result and the black solid line represents the resummed result at NLL. 
The blue solid line is the NLL resummed result when we consider the electronic jet (with the radius $r=0.1$) to include collinear photons around the electron. The error bands for both of the resummed results have been estimated in the same way as FIG.~\ref{fig1}. 
} 
\end{figure}

In FIG.~\ref{fig2}, we show the resummed result for the asymmetric contribution in the leptonic $B$ decays, $\Delta^B_{\mu e}$, in the range  $\ec \in [10,50]~\mr{MeV}$, and compare it with the fixed-order NLO result.    
Like the exclusive $Z$ decays, the asymmetric contribution is fairly large. For two choices of $\ec$, the NLL resummed results are
\bea 
\Delta^B_{\mu e} &=& 0.083\pm 0.005~~(\ec = 20~\mr{MeV}), \\
\Delta^B_{\mu e} &=& 0.064\pm 0.005~~(\ec = 40~\mr{MeV}). 
\eea 
Neglecting $\mc{O}(m_{\tau}^2/m_B^2)$ in QED corrections, the asymmetric contribution between the decays with $\tau$ and with $\mu$ is estimated as 
\bea 
\Delta^B_{\tau\mu} &=& 0.031\pm 0.004~~(\ec = 20~\mr{MeV}), \\
\Delta^B_{\tau\mu} &=& 0.028\pm 0.004~~(\ec = 40~\mr{MeV}). 
\eea 
 
Analogously, using the resummed result we can estimate the ratio between the decays, $K^-\to l^-\bar{\nu}_l~(l=\mu,e)$, where QED corrections of $\mc{O}(m_{\mu}^2/m_K^2)$ may safely be ignored. For the choices $\ec = 10~\mr{MeV}$ and $20~\mr{MeV}$ in the kaon rest frame,  
the ratios $R^K_{e\mu}~(=(R^K_{\mu e})^{-1})$ are respectively    
\bea 
R^K_{e\mu} &=& (2.455\pm 0.010)\times 10^{-5}, \\
R^K_{e\mu} &=& (2.490\pm 0.010)\times 10^{-5},
\eea
where the asymmetric contributions from the resummed results are 
\bea
\Delta^K_{e\mu} = -(4.00\pm 0.43)\%~~(\ec = 10~\mr{MeV}), \\
\Delta^K_{e\mu} = -(2.54\pm 0.41)\%~~(\ec = 20~\mr{MeV}).
\eea
These results are compatible with previously analyzed theoretical~\cite{Cirigliano:2007xi,Cirigliano:2007ga} and experimental~\cite{Lazzeroni:2012cx} results. 

When we have an electron in the final state of the exclusive process, it might be experimentally difficult to disentangle collinear photons from the electron for the isolation. 
So, rather than strict exclusive events, we may consider more inclusive events that include photons within a cone with a small radius around the electron. In this case, for the electronic sector, the collinear and the csoft functions in Eqs.~\eqref{rZ} and \eqref{RB12} are modified and the NLO contributions are~\cite{Kim:2020dgu,Dai:2018ywt}
\begin{widetext}
\bea
C_r^{(1)} (E_jr,m,\mu) &=&  \frac{\alpha}{2\pi} \Bigl[ \frac{3+b}{2(1+b)} \ln\frac{\mu^2}{E_j^2 r^2+m^2} +\frac{1}{2} \ln^2 \frac{\mu^2}{E_j^2 r^2+m^2}+ \frac{2}{1+b}+f(b)+g(b) + h(b) + 2 -\frac{\pi^2}{12} \Bigr],
\nnb \\
\label{CBr}
B_r^{(1)}(\beta, E_jr,m,\mu) &=& \frac{\alpha}{2\pi} \Bigl[\frac{b}{1+b} \ln\frac{\mu^2}{4\beta^2(E_j^2 r^2+m^2)}-\frac{1}{2} \ln^2\frac{\mu^2}{4\beta^2(E_j^2 r^2+m^2)}   - h(b) + \frac{\pi^2}{12} \Bigr],
\eea
\end{widetext}
where $E_j$ is the energy of ``the electronic jet'' including collinear photons, $r$ is the jet radius, and $b \equiv m^2/(E_j^2r^2)$. The functions $f(b)$, $g(b)$, and $h(b)$ are given by  
\bea
f(b) &=& \int^1_0 dz \frac{1+z^2}{1-z} \ln\frac{z^2+b}{1+b}, 
\nnb \\
\label{fgh}
g(b) &=& \int^1_0 dz \frac{2z}{1-z}\Bigl(\frac{1}{1+b}-\frac{z^2}{z^2+b}\Bigr), \\
h(b) &=& \frac{\ln(1+b)}{1+b} - \frac{1}{2} \ln^2 (1+b) - \mr{Li}_2 (-b). \nnb 
\eea
In the limit $r\to 0~(b\to \infty)$, these functions become $f(\infty) = g(\infty) =0$ and $h(\infty) = \pi^2/6$. Hence Eq.~\eqref{CBr} in this limit recovers the results in Eq.~\eqref{fNLO}.  

From Eq.~\eqref{CBr}, the characteristic scales for $C_r$ and $B_r$ are 
\be
\mu_{C} \sim \sqrt{E_j^2r^2+m^2},~~\mu_{B} \sim 2\beta\sqrt{E_j^2r^2+m^2}. 
\ee
Employing $C_r$ and $B_r$ for the process $B^-\to e^-\bar{\nu}_e$, we resum the large logarithms in the ratio $R^B_{\mu e}$ by evolving the factorized functions. 
For our numerical analysis, we set $E_j = m_b/2= 2.4~\mr{GeV}$ and the cone radius for the electronic jet as $r=0.1$ in the $B$ meson rest frame. In FIG.~\ref{fig2} we illustrate the resummed result for the asymmetric contribution as a blue solid line. 
Since the electronic jet takes larger phase space, the decay width for $B^-\to e^-\bar{\nu}_e$ is enhanced and the asymmetric contribution becomes $\Delta^B_{\mu e}\sim 0$. In this case the characteristic scale $\mu_{C}$ for the electronic jet is given by $\mu_{C} \sim E_j r = 0.24~\mr{GeV}$, which is numerically close to the scale for the muon, $\mu_C \sim m_{\mu}$. So in this situation the dependence on the large logarithms in the ratio roughly cancel.

In summary, we find that the SM can predict fairly large lepton universality violations in high energy exclusive processes. The energetic leptonic sector in the process decouples and involves large logarithms of the lepton mass and the photon energy cut. Using the factorization result, we can systematically resum large logarithms in the ratio between different lepton flavors. The result, Eq.~\eqref{resR}, can be universally applied to different processes including heavy particle decays to a lepton (pair). 
Our analysis can be also extended to Drell-Yan process near threshold and  semileptonic $B$ decays when leptons have a large invariant mass $(q)$. For example, our result $(\sim R_{\mu e}^Z|_{m_Z \to m_B})$
might be roughly comparable with the experimental data for $B\to K^{(*)} l^+l^-$ with $q^2 > 10.11~\mr{GeV}^2$~\cite{Lees:2012tva}
even though we need to do more refined analysis on the three body decay.\footnote{
Our factorization is applicable when the energetic lepton pair move in opposite directions in the $B$ rest frame, i.e., in the large $q^2 (\sim m_B^2 - m_K^2)$ region. We thus cannot directly compare to the recent LHCb experimental result~\cite{Aaij:2021vac}, which  investigated the region $1.1~\mr{GeV}^2 < q^2 < 6~\mr{GeV}^2$.   
}  

Because we deal with  ideal exclusive processes in this paper, the theoretical analysis is clear. 
However, experimentally analyzing the same situations would be more challenging. 
It is possible that the realistic, complicated experimental environment would introduce additional IR physics. 
This can affect the result of the lepton universality violation compared to the strict exclusive processes, 
as demonstrated with the large difference on whether the electron in the final state is isolated or not.   

LD was supported by the Foreign Postdoctoral Fellowship Program of the Israel Academy of Sciences and Humanities, Israeli Science Foundation (ISF) grant \#1635/16, and Binational Science Foundation grant  \#2018722. This work has been performed in the framework of COST Action CA 15213 ``Theory of hot matter and relativistic heavy-ion collisions" (THOR), MSCA RISE 823947 ``Heavy ion collisions: collectivity and precision in saturation physics''  (HI\-EIC) and has received funding from the European Un\-ion's Horizon 2020 research and innovation programm under grant agreement No. 824093. CK was supported by Basic Science Research Program through the 
National Research Foundation of Korea (NRF) funded by the Ministry of Science and ICT (Grant No. NRF-2017R1A2B4010511). AL is supported in part by the National Science Foundation under Grant No. PHY-1820760. 



\end{document}